\documentclass{article} 
\usepackage{iclr2019_conference,times}

\usepackage{url}
\usepackage[utf8]{inputenc} 
\usepackage[T1]{fontenc}    
\usepackage{booktabs}       
\usepackage{amsfonts}       
\usepackage{nicefrac}       
\usepackage{microtype}      
\usepackage{amsmath}
\usepackage{geometry}
\usepackage{amssymb}
\usepackage{graphicx}
\usepackage{caption}
\usepackage{subcaption}
\usepackage{amsthm}
\usepackage{bbm}
\usepackage{booktabs}
\usepackage{array,multirow}
\usepackage{scrextend}
\usepackage{enumitem}
\setlist[description]{style=multiline,leftmargin=4cm}

\usepackage[frozencache,cachedir=minted]{minted}
\setminted[python]{
  framesep=2mm,
  baselinestretch=1.2,  
  fontsize=\footnotesize}
\setmintedinline[python]{bgcolor=white}
\renewcommand{\DeleteFile}[2][]{}  

\usepackage{xcolor}
\usepackage[title]{appendix}
\usepackage{listings}
\usepackage{setspace}

\usepackage[english]{babel}
\usepackage{mdframed}   

\usepackage[hidelinks]{hyperref}

\usepackage{amsmath,amsfonts,bm}

\def\eqref#1{(\ref{#1})}

\def\1{\bm{1}}

\def\rvc{{\mathbf{c}}}

\def\rvw{{\mathbf{w}}}

\DeclareMathAlphabet{\mathsfit}{\encodingdefault}{\sfdefault}{m}{sl}
\SetMathAlphabet{\mathsfit}{bold}{\encodingdefault}{\sfdefault}{bx}{n}

\newmdtheoremenv[%
linecolor=gray,leftmargin=2,%
rightmargin=2,
backgroundcolor=gray!40,%
innertopmargin=5pt,%
ntheorem]{ex}{Example}[section]

\definecolor{light-gray}{gray}{0.95}

\newcommand{\todo}[1]{\textcolor{red}{TODO: #1}}

\renewcommand{\todo}[1]{} 

\newcommand{\email}[1]{\href{mailto:#1}{\nolinkurl{#1}}}

\usepackage{pdftexcmds}  
\usepackage[dont-mess-around]{fnpct}
\usepackage{etoolbox}
\makeatletter
\patchcmd\maketitle{\setcounter{footnote}{0}}{}{}{}
\patchcmd\maketitle{%
 \renewcommand\thefootnote{\@fnsymbol\c@footnote}}{\AdaptNote\thanks\multthanks}{}{}
\patchcmd\maketitle{%
 \def\@makefnmark{\rlap{\@textsuperscript{\normalfont\@thefnmark}}}}{}{}{}
\makeatother

\title{
  A Python Library for Empirical Calibration
}

\author{Xiaojing Wang\\
 Google\\
 \texttt{xiaojingw@google.com}
 \And
 Jingang Miao\\
 Google\\
 \texttt{jmiao@google.com}
 \And
 Yunting Sun\\
 Google\\
 \texttt{ytsun@google.com}
}
\usepackage[compact]{titlesec}
\titlespacing{\section}{0pt}{2ex}{1ex}
\titlespacing{\subsection}{0pt}{1ex}{0ex}
\titlespacing{\subsubsection}{0pt}{0.5ex}{0ex}

\newcommand{\pkg}[1]{{\fontseries{b}\selectfont #1}}

\iclrfinalcopy 
\DeclareUnicodeCharacter{2212}{-}
\begin{document}

\maketitle

\begin{abstract}
Dealing with biased data samples is a common task across many statistical
fields. In survey sampling, bias often occurs due to unrepresentative
samples. In causal studies with observational data, the treated versus untreated
group assignment is often correlated with covariates, i.e., not random.
Empirical calibration is a generic weighting method that presents a unified view
on correcting or reducing the data biases for the tasks mentioned above. We
provide a Python library \pkg{EC} to compute the empirical calibration
weights. The problem is formulated as convex optimization and solved
efficiently in the dual form. Compared to existing software, \pkg{EC} is both
more efficient and robust. \pkg{EC} also accommodates different optimization
objectives, supports weight clipping, and allows inexact calibration, which
improves usability. We demonstrate its usage across various experiments with
both simulated and real-world data.
\end{abstract}

\section{Introduction}

Biased data samples are prevalent in statistical problems. In survey sampling,
unrepresentative data can occur when certain sub-population are under- or
over-represented in the sample. For example, researchers may decide to
over-sample females in a marketing study of cosmetics using differential
sampling probabilities, which, if not accounted for, would lead to selection
bias. Further, once invited to join the study with a cash incentive, richer
individuals may be less likely to participate, and the differential response
rates may cause self-selection bias. To guard against such biases, sampling
weights, non-response weights, and calibration weights are used at different
stages of survey data adjustment. In causal inference with observational data,
the treated versus untreated group assignment is often correlated with covariates,
i.e., not random. Weighting can be applied to untreated individuals such that
the reweighted untreated group better mimics the control group as in a
randomized control experiment.

Weighting is a generic bias correction technique that can take different
forms. One is inverse probability weighting, where the inverse of
\emph{probability/propensity} of being included in the sample or receiving the
treatment is used as weights. Such probability is either known or estimated from
a propensity model. Popular examples include the Horvitz-Thompson estimator
\citep{horvitz1952generalization} in survey sampling and inverse propensity
score weighting \citep{rosenbaum1983central} in causal inference.

Empirical calibration takes a different form, where instead of trying to
summarize the systematic differences with one \emph{probability/propensity}, one
seeks weights that directly balance out the covariates but deviate least from
the uniform weights to reduce the inflation of variance.

Solving the empirical calibration weights can be formulated as a convex
optimization problem. Suppose there are $n$ biased data points
$\{X_i\}_{i=1}^n$, and $p$ marginal constraints $\{\bar c_j\}_{j=1}^p$. The goal
is to find a $n$-dimensional weight vector $\rvw$ that is the least unequal
while subjecting to the weight normalization and marginal balance constraints.
\begin{align*}
\underset{\rvw}{\text{minimize}} \quad & L(\rvw)\\
\text{subject to} \quad
& \sum_{i=1}^n w_i c_j(X_i) = \bar c_j, \quad j = 1, \ldots, p, \\
& \sum_{i=1}^n w_i = 1, \\
& w_i \ge 0, \quad i = 1, \ldots, n.
\end{align*}
where $L$ is the convex loss function and $c_j(X_i)$ is the $j$-th
transformation.

There are variations of the unequalness definition, which determines the loss
function $L$. \citet{hainmueller2012entropy} proposed \emph{entropy balancing}
in the context of causal inference with observational data. It corresponds to
the entropy loss
\begin{equation*}
  L(\rvw) = \sum_{i=1}^n w_i \log (w_i),
\end{equation*}
which can be viewed as the Kullback-Leibler divergence between $\rvw$ and
uniform weights. Alternatively, \citep{deville1992calibration} used the
Euclidean or squared distance between $\rvw$ and some base weights --- typically
uniform weights. It is equivalent to a quadratic loss
\begin{equation*}
  L(\rvw) = \frac{1}{2} \sum_{i=1}^n w_i^2.
\end{equation*}
Further, minimizing this quadratic loss can be seen as maximizing the effective
sample size \citep{kish1965survey}:
\begin{equation*}
\frac{(\sum_{i=1}^n w_i)^2}{\sum_{i=1}^n w_i^2}.
\end{equation*}
More broadly, both entropy and quadratic losses can be considered as special
cases of the empirical likelihood \citep{owen2001empirical}.

We provide a Python library \pkg{EC} that supports both entropy and quadratic
loss functions. When there is no feasible solution to satisfy the balance
constraints, \pkg{EC} allows for inexact constraints, which improves the
robustness and usability. For some applications, it can be desirable to bound
the weights to a certain range, \pkg{EC} conveniently has the weight restraint
option built-in, which is superior to post optimization clipping or
winsorization.

This paper is structured as follows. Section 2 briefly presents \pkg{EC}'s
interface. In section 3, we give a short tour of various applications of
empirical calibration, and present experimental results on both simulated and
real-world data. Some implementation issues are discussed in section 4.
Concluding remarks are given in section 5.

\section{Software}

The \pkg{EC} library is available at
\url{https://github.com/google/empirical_calibration} and can be imported as
\begin{minted}{python}
import empirical_calibration as ec
\end{minted}

The primary interface is function \mintinline{python}{ec.calibrate}:

\begin{minted}{python}
def calibrate(covariates: np.ndarray,
              target_covariates: np.ndarray,
              target_weights: np.ndarray = None,
              autoscale: bool = False,
              objective: Objective = ec.Objective.ENTROPY,
              max_weight: float = 1.0,
              l2_norm: float = 0) -> Tuple[np.ndarray, bool]:
\end{minted}

Its arguments are
\begin{description}
\item[\mintinline{python}{covariates}] $X$, covariates to be calibrated. All
  values must be numeric. For categorical values, the
  \mintinline{python}{from_formula} function is often more convenient.
\item[\mintinline{python}{target_covariates}] covariates to be used as target
  in calibration. The sum of each column is a marginal constraint ($\bar{c}_j$).
  The number of columns should match \mintinline{python}{covariates}.
\item[\mintinline{python}{target_weights}] weights for
  \mintinline{python}{target_covariates}. These are needed when the
  \mintinline{python}{target_covariates} themselves have weights.
  Its length must equal the number of rows in
  \mintinline{python}{target_covariates}.
  If None, equal weights are used.
\item[\mintinline{python}{autoscale}] whether to scale covariates to [0, 1]
  and apply the same scaling to \mintinline{python}{target covariates}.
  Setting it to True might help improve numerical stability.
\item[\mintinline{python}{objective}] the objective of the convex optimization
  problem. Supported values are
  \mintinline{python}{ec.Objective.ENTROPY} ($\sum_{i=1}^n w_i \log (w_i)$) and
  \mintinline{python}{ec.Objective.QUADRATIC}
  ($\frac{1}{2} \sum_{i=1}^n w_i^2$).
\item[\mintinline{python}{max_weight}] the upper bound on weights. Must be
  between uniform weight (1 / number of rows in \mintinline{python}{covariates})
  and 1.0.
\item[\mintinline{python}{l2_norm}] $\epsilon$, the L2 norm of the covariates balance
  constraint; i.e., the Euclidean distance between the weighted mean of
  covariates and the simple mean of target covariates after balancing.
\end{description}

It returns a tuple \mintinline{python}{(weights, success)}, where
\begin{description}
\item[\mintinline{python}{weights}] The weights for the  subjects. They should
  sum up to 1.
\item[\mintinline{python}{success}] Whether the constrained optimization
  succeeds.
\end{description}

Entropy objective is a common choice for causal inference problems, while the
quadratic objective is more widely used in the context of survey calibration.
One practical difference is that the entropy objective yields strictly positive
weights while the quadratic objective allows zero weights where zero-weight data
points are effectively discarded in subsequent analysis.

The given constraints may yield no feasible solution, in which
case one can either drop some constraints or soften the hard constraints by
specifying a maximal distance allowed between the weighted covariates and the
target covariates via the \mintinline{python}{l2_norm} argument. Further,
instead of manually specifying the soft constraint, one can use
\mintinline{python}{ec.maybe_exact_calibrate}, which automatically chooses the
smallest soft margin that yields a feasible solution.

Also provided is the \mintinline{python}{from_formula} interface, which is
parallel to the functions above with \mintinline{python}{covariates} and
\mintinline{python}{target_covariates} replaced with

\begin{description}
\item[\mintinline{python}{formula}] Formula used to generate design matrix.
      No outcome variable allowed.
\item[\mintinline{python}{df}] Data to be calibrated.
\item[\mintinline{python}{target_df}] Data containing the target.
\end{description}

\mintinline{python}{formula} here follows \pkg{patsy}-style so that one can
conveniently construct square terms, interaction terms, etc. For example,
\mintinline{python}{formula='~ -1 + x1 + x2 ** 2 + x3:x4'} tells \pkg{EC} to
match the first moment of $x1$, the second moment of $x2$ as well as the
interaction between $x3$ and $x4$. This saves the user the trouble of
manually constructing these quantities.

We illustrate the usage of these interfaces in the Applications section below.

\section{Applications}

\subsection{Survey Calibration}

Surveys or samples are drawn to make inferences about the target population,
which is the bread and butter of statistics. A valid inference is only possible when
the survey or sample is representative of the target population.

Survey weighting is commonly used to correct for unequal sampling probability,
differential response rates, and coverage errors \citep{deville1992calibration,
  kott2006using}. Calibration, which is often one of the steps in survey
weighting, makes use of information about the target population that is
available \emph{after} the sample is collected (auxiliary information). For
example, for a survey of internet users, such information may be found in the
Current Population Survey Computer and Internet Use Supplement
\citep{ryan2017computer}, including distributions of internet users' gender,
age, education, and household income. Calibration aims to make the sample more
representative of the target population. Popular methods of using auxiliary
information include ratio estimation \citep{fuller2011sampling},
post-stratification or raking \citep{little1993post}, and calibration estimation
\citep{deville1992calibration}.

Weighted samples can be used to estimate the population average of survey
response. The weights are often constructed such that weighted moments ---
typically the first two moments --- agree with known population benchmarks. The
benchmarks could come from census data or other large scale and high-quality
surveys. Meanwhile, the calibration weights are chosen to least deviate from the
base weights, which account for sampling design and differential
non-response. In the simplest form, the base weights are uniform.

\subsubsection{Estimate population mean}

A common use case of surveys is to estimate the mean or total of a quantity. We
replicate the \emph{direct standardization} \citep{fleiss2003} example by
\cite{fu2018blog}. Data
is obtained from the \pkg{CVXR} package \citep{fu2017cvxr}, where
\mintinline{python}{dspop} contains 1000 rows with columns $\text{sex} \sim
\text{Bernoulli}(0.5)$, $\text{age} \sim \text{Uniform}(10, 60)$, and $y_i \sim
N(5 \times \text{sex}_i + 0.1 \times \text{age}_i,
1)$. \mintinline{python}{dssamp} contains a skewed sample of 100 rows with small
values of $y$ over-represented, thus biasing its distribution downwards.

\begin{minted}{python}
cols = ['sex', 'age']
weights, _ = ec.maybe_exact_calibrate(
    covariates=dssamp[cols],
    target_covariates=dspop[cols],
    objective=ec.Objective.ENTROPY
    )

print('True mean of y: {}'.format(dspop['y'].mean()))
print('Unweighted sample mean: {}'.format(dssamp['y'].mean()))
print('Weighted mean: {}'.format(np.average(dssamp['y'], weights=weights)))
\end{minted}

The true mean of $y$ based on the data generating process is 6.0.
Using the generated population of size 1000 and a sample of size 100
contained in the \pkg{CVXR} package, the population mean is 6.01, but the
mean of the skewed sample is 3.76, which is a gross underestimation.
With empirical calibration, however, the weighted mean
is 5.82, which is closer to the population mean.

\begin{figure}[tbp]
  \center
    \includegraphics[height = 5cm, width = 0.5\textwidth]{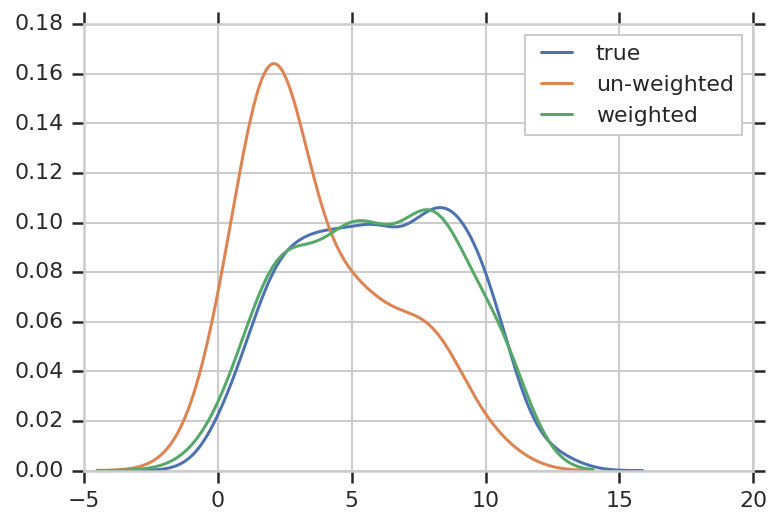}
  \caption{Estimate population mean: Kernel density estimates of the true, unweighted, and weighted $y$.}
  \label{fig:direct_standardization}
\end{figure}

The kernel density plots (Figure \ref{fig:direct_standardization}) show that the
unweighted curve is biased toward smaller values of $y$, but the weights help
recover the true density.

\subsection{Observational Data with a Binary Treatment}


Empirical calibration is also applicable in observational studies with a binary
treatment. It aims to achieve covariate balance by calibrating the untreated
group against the treated group. It assigns non-negative weights to individual
control units such that certain moments --- typically means --- of covariates
between the treatment group and the untreated group are matched. The weighted
untreated units then mimic a control sample as if it was from a randomized
experiment.

We follow the potential outcome language of Rubin causal model to describe the
inference problem with a binary treatment. Suppose each unit $i$ is associated
with a pair of potential outcomes: $Y_i(1)$ if treated ($T_i = 1$) and $Y_i(0)$
if untreated ($T_i = 0$). The treatment effect for this unit is defined as
$\tau_i = Y_i(1) - Y_i(0)$. Inference of $\tau_i$'s becomes a missing data
problem since the two potential outcomes are never both observed --- commonly
known as the \emph{fundamental problem of causal inference}. If $Y_i(T_i)$ is
observed, $Y_i(1 - T_i)$ is the \emph{counterfactual}, i.e., what would have
been observed if $T_i$ did not occur.

We limit the discussion to the sample \emph{average treatment effect on the
  treated} (ATT) estimand:
\begin{equation}\label{eq:satt}
\frac{1}{n_1}\sum_{i: T_i=1} [Y_i(1) - Y_i(0)],
\end{equation}
where $n_1$ is the number of treated units. The first term in~\eqref{eq:satt} is
straightforward to compute using the treated observations. The problem boils
down to estimating the second term --- the mean of unobserved counterfactuals
--- from the control observations.

In \textbf{experimental} settings, the treatment assignment is independent of
the outcome, $Y(1), Y(0) \perp T$, and one can simply use $\frac{1}{n_0}
\sum_{i: T_i=0}Y_i(0)$ as an estimate of $\frac{1}{n_1} \sum_{i: T_i=1} Y_i(0)$.
In \textbf{observational} studies, however, due to the treatment selection bias,
the treated group is often systematically different from the control group,
rendering the latter two quantities unequal. Conventionally, we assume observed
covariates $X$ contain all the information about the selection bias (i.e., there
is no unobserved confounding variable, which is a strong assumption), and
estimate ATT in two stages: 1) ``design'' purely based on matching of observed
covariates $X$ and 2) outcome analysis of $Y$. Stage 1 equates or balances the
distributions of covariates between the treated and control groups. Matching and
weighting are the two main approaches to achieve the balance, see
Appendix~\ref{app:causal} for a short tour of common methods. Stage 2 compares
the outcomes of the treated and control units, and estimates the causal effect
of the treatment. It generally involves some regression adjustments to account
for the small residual covariate imbalance between the groups after stage 1. The
outcome regression must be able to accept weights if stage 1 is done via
weighting. The matching methods and outcome analysis in the two stages have been
shown to work best in combination \citep{rubin1973use}. The intuition is the
same as that behind the \emph{double-robust} estimators
\citep{robins1994estimation}, which are asymptotically unbiased if either
the propensity score matching model or the outcome regression model is correctly
specified.

For empirical calibration weighting, we focus on the \textbf{affine estimator}:
\begin{equation}
\hat \tau_{\text{EC}} = \frac{1}{n_1} \sum_{i: T_i=1} Y_i - \sum_{i: T_i=0} w_i Y_i,
\end{equation}
where $n_1$ is the number of treatment units and $w_i$'s are the empirical
calibration weights that sum up to 1. The first term is the simple average of
the observed outcomes for the treated units, and the second term is the weighted
average of the observed outcomes for the control units.

\subsubsection{Kang-Schafer Simulation}

\citet{kang2007demystifying} used a simulation study to illustrate the selection
bias of outcome under informative non-response. The study became a standard benchmark
to compare different estimators for causal estimands.

The true set of covariates is generated independently and identically
distributed from the standard normal distribution
\begin{equation*}
(Z_1, Z_2, Z_3, Z_4) \sim N(0, \mathbf{I}_4).
\end{equation*}
The outcome is generated as
\begin{equation*}
Y = 210 + 27.4 Z_1 + 13.7 Z_2 + 13.7 Z_3 + 13.7 Z_4 + \epsilon,
\end{equation*}
where $\epsilon \sim N(0, 1)$.

The propensity score is defined as
\[
Pr(T = 1 | Z) = \text{expit}(-Z_1 + 0.5 Z_2 - 0.25 Z_3 - 0.1 Z_4).
\]

This mechanism produces an equal-sized treated and control group
on average. Given the covariates, the outcome is independent of the treatment
assignment; thus the true ATT is zero. The overall outcome mean is 210. Due to
the treatment selection bias, the outcome mean for the treated group (200) is
lower than that of the control group (220).

A typical exercise is to examine the performance of an observational method
under both correctly specified and misspecified propensity score and/or outcome
regression models. Misspecification occurs when the following nonlinear
transformation $X_i$'s are observed in place of the true covariates
\begin{align*}
X_{i1} & = \exp(Z_{i1}/2), \\
X_{i2} & = Z_{i2} / (1 + \exp(Z_{i1})) + 10, \\
X_{i3} & = (Z_{i1} Z_{i3} / 25 + 0.6)^3, \\
X_{i4} & = (Z_{i2} + Z_{i4} + 20)^2.
\end{align*}

To estimate ATT, we use empirical calibration to weight the control group so
that it matches the treatment group in terms of their covariate distribution.
Simulations with a sample of 1000 are used.

Without weights, the bias is 20.2, and RMSE is 20.3.  With correctly specified
covariates to match ($Z1, \dots, Z4$), the bias is almost zero (0.001), and RMSE
is 0.09, With incorrectly specified covariates to match ($X1, \dots, X4$), bias
is -4.5, and RMSE is 4.6.  With matching on additional transformations of
incorrectly specified covariates ($X1, \dots, X4$ plus interactions terms and
log-transformed versions), bias is -1.8 and RMSE is 2.0. It is evident that
weighting reduces the bias of the ATT estimate significantly.

We also use  empirical calibration to estimate the population
mean and compare it with the results reported in
\citet{kang2007demystifying}.
The treatment group is weighted so that it matches
the population in terms of their covariate distribution.
The estimator is the weighted value of $y$ in the treatment group. Again,
simulations with a sample of 1000 are used.
With correctly specified covariates to match ($Z1, \dots, Z4$),
the bias is almost zero (0.0003),
and RMSE is 1.13, which is better than all the methods described in
\citet{kang2007demystifying}.

\subsubsection{LaLonde Data Analysis}

The \citet{lalonde1986evaluating} data is another canonical benchmark in the
causal inference literature. It consists of three groups for evaluating the
effect of a large scale job training program --- the National Supported Work
Demonstration (NSW). An experimental treatment group with 185 observations, an
experimental control group with 260 observations, and an observational control
group was drawn from the Current Population Survey (CPS), with 15,992 observations.
The outcome variable was the post-intervention earnings in 1978, and common
covariates for all three datasets include age, education, earnings in 1974,
earnings in 1975, and binary indicators of being black, being Hispanic, being
married, and having completed high school.

Using the experimental control, the simple difference-in-means estimate of the
average treatment effect on the treated is 1794.3 with a 95\% confidence
interval of [479.2, 3109.5]. A linear regression with earning in 1978 as the
outcome and treatment or control indicator together with other variables as
independent variables gives an estimate of 1698 with a 95\% confidence interval
of [458, 2938], which is narrower than the difference in means estimate.

We apply entropy balancing on the observational control with respect to 52
covariates as described in \citet{hainmueller2012entropy}, and the difference in
means estimate is 1571, which is identical to what was reported in
\citet{hainmueller2012entropy}. Using quadratic balancing, the point estimate is
1713, which is very close to the estimate using the experimental control.

\subsection{Geo Experiments}

An advertiser must measure the impact and return on investment
of its online ad campaigns. Randomized controlled experiment or A/B testing is
considered the gold standard for drawing causal conclusions. A user-level
experiment randomly assigns users to the treatment or control group. No ad is
served to the control users. The causal effect of advertising can then be
estimated by comparing the observed outcomes between the two groups. It can be
difficult however to maintain the integrity of the user level group assignment
due to various issues such as user signed in and signed out, cookie churn,
cross-device usage.

Geo-level experiments offer an attractive alternative to user-level experiments
by experimenting at the geographic level. A region is first partitioned into
non-overlapping subregions, or simply ``geos'', which are then randomly assigned
to a control or treatment condition. Each geo realizes its assigned treatment
condition through the use of geo-targeted advertising. For example, an online ad
campaign is run in the treatment geos but not in the control geos. The behavior
changes, such as website visits and purchases, of customers in the treatment
geos can then be attributed to the ad campaign.

Despite the random assignment of treatment and control geos, the systematic
difference between the two groups can still be substantial, because there may only be
a small number of highly heterogeneous geos available for experimentation. A
common solution is first to pair similar geo together and then apply random
assignment within each pair. The issue is that some geo, for example, New York
City, is too large to find a matching geo and cannot be sufficiently balanced
out.

Empirical calibration can help lessen the imbalance of the geo-level
randomization, and thus reduce the bias of the causal effect estimate. It finds
weights for the control geos such that the weighted average of the control geo
time series matches that of the treatment geos in the pretest period. The same
weights are then applied to control time series in the test period as the
treatment counterfactual time series.

Empirical calibration does not aggregate time series across pretest or test
period as in geo-based regression (GBR) \citep{vaver2011measuring}, nor does it
aggregate data across control geos as in the time-based regression (TBR)
\citep{kerman2017estimating}. It leverages individual time series at each geo
for causal inference. Although GBR allows geo level weights in its linear model
to account for heteroscedasticity caused by the differences in geo size, unlike
in empirical calibration, GBR's weights are determined in a less principled way
--- simply the inverse of the pretest period metrics. Empirical calibration
shares the spirit of using contemporaneous controls with Bayesian structural
time series \citep{brodersen2015inferring} models but does not need to specify
a time series model, thus is more robust to model misspecification. We
demonstrate the usage of empirical calibration for geo experiments using the
data found in \citet{geox2017package}. The goal is to estimate the treatment
effect on geo 30.

\begin{figure}[tbp]
  \center
  \begin{subfigure}{.48\textwidth}
    \caption{Unweighted mean of controls as counterfactual}
    \includegraphics[height = 5cm, width = \textwidth]{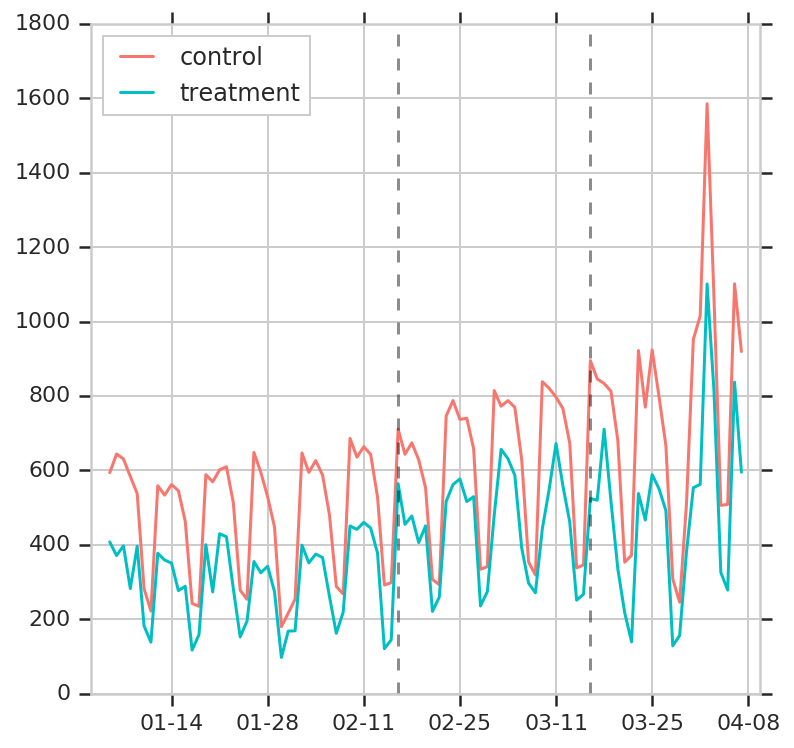}
  \end{subfigure}
  \begin{subfigure}{.48\textwidth}
    \caption{Weighted mean of controls as counterfactual}
    \includegraphics[height = 5cm, width = \textwidth]{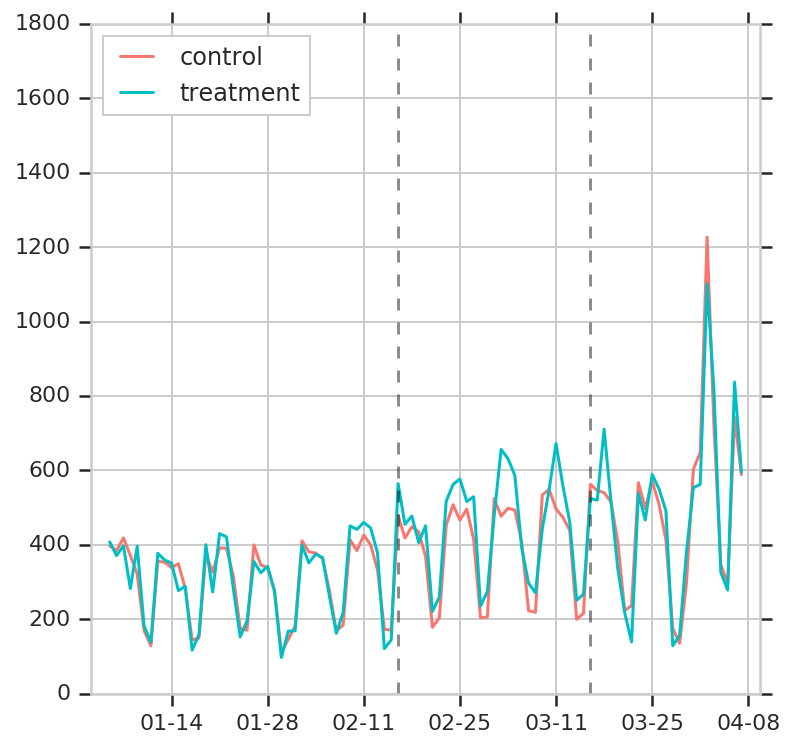}
  \end{subfigure}
  \caption{Causal inference for geo 30 with unweighted and weighted mean of
     control geos as the counterfactual estimate. The pretest period is to the left
     of the dashed lines; the intervention period is between the dashed lines, and
     cooldown period is to the right of the dashed lines.}
  \label{fig:geox}
\end{figure}

The unweighted mean of controls is systematically above that of the treatment
geo in the pretest period, which suggests that the difference between the
unweighted means in the test period is not an indication of any causal effect of
the treatment. The weighted mean of controls, on the other hand, matches the
treatment geo much better in the pretest period, which leads to a more
reasonable counterfactual estimate.

Instead of looking at one treatment geo at a time, one can also choose to study
the mean of all treatment geos as compared to control geos.



\section{Implementation}

Stata package \pkg{Ebalance} \citep{hainmueller2013ebalance} and R package
\href{https://cran.r-project.org/web/packages/ebal/index.html}{\pkg{ebal}}
\citep{hainmueller2014ebal} package finds the entropy balancing weights using
the algorithm proposed by \citet{hainmueller2012entropy}. It exploits the dual
form of convex optimization, where the primal weights can be expressed as a
log-linear function of the covariates specified in the moment conditions. The
dual problem is more tractable than the primal problem because it is unconstrained and the
dimensionality reduces to the number of marginal constraints. A
Levenberg–Marquardt scheme is employed to update the dual solution iteratively.
When the problem is feasible, the algorithm is globally convergent.

R package \pkg{survey} offers some overlapping functionality with our python
implementation. Calling \mintinline{R}{survey::calibrate} with the argument
\mintinline{R}{calfun="linear"} is equivalent to using
\mintinline{python}{ec.calibrate} with a quadratic objective. If no upper or
lower bounds are placed on the weights, a closed-form solution is used; however,
negative weights are possible. If, on the other hand, bounds are set, then an
iterative algorithm is used instead.  One advantage of our implementation is
that \mintinline{R}{survey::calibrate} would fail when there is no feasible
solution, while \mintinline{python}{ec.calibration} allows gradual relaxation of
the constraints until a feasible solution can be found.

Alternatively, both the entropy balancing and quadratic balancing weights can be
solved by general purpose convex optimization software. Python library
\pkg{CVXPY} \citep{diamond2016cvxpy} and R package \pkg{CVXR} \citep{fu2017cvxr}
are two common choices, which provide domain-specific modeling language and
interface with open-source solvers such as \pkg{ECOS} \citep{domahidi2013ecos}
and \pkg{SCS} \citep{o2016conic}.

Our implementation mostly follows the entropy balancing dual solution. We also
express the primal weights as a function of the covariates and Lagrangian
multipliers. The function is log-linear for entropy balancing and linear with a
ReLU activation for quadratic balancing. Instead of relying on
Levenberg–Marquardt update, we find the dual solution by solving a set of
non-linear equations with Python's \mintinline{python}{scipy.optimize.fsolve},
see Appendix~\ref{app:cvx} for details.

There could be no feasible solution when the sample size is small, or the number
of imposing constraints is large. In the context of observational studies with a
binary treatment, it is particularly common to have no feasible solution when
there was limited overlap between the treated and untreated covariate
distributions. Instead of forcing our users to drop certain marginal constraints
altogether, we provide an option for our users to relax the hard constraints. In
our experience, a small constraint relaxation often leads to feasible solutions
without too much degradation on the weights, significantly improving the numerical
computation experience.

In other applications, it might be desirable to restrict the weights to a
particular range. The conventional solution is to apply weight clipping or
winsorization in a post-processing step, but it leads to a violation of the weight
normalization constraint and sub-optimal weight solutions. Our implementation
allows the weight clipping imposed directly as an additional constraint in the
optimization problem, yielding optimal weights satisfying all constraints at
once.

\subsection{Benchmark against R Packages \pkg{ebal} and \pkg{CVXR}}

Benchmarking was done against R package \pkg{ebal} version 0.1-6 and package
\pkg{CVXR} version 0.99-4 on a \citet{kang2007demystifying} simulation of sample
size 2,000. Computation was done for 200 times for each package. Package
\pkg{EC} was 22 and 73 times as fast as \pkg{ebal} and \pkg{CVXR} respectively
(Table \ref{table:benchmark}). Further, Package \pkg{EC} had less variability in the
performance times.

\begin{table}[ht]
\centering
\begin{tabular}[t]{lrrr}
  \toprule
  Package & Mean & Min & Max \\
  \midrule
  \pkg{ebal} & 30.7 & 26.3 & 121.7 \\
  \pkg{CVXR} & 102.2 & 88.4 & 140.6 \\
  \pkg{EC} & 1.4 & 1.3 & 2.1 \\
  \bottomrule
\end{tabular}
\caption{Performance benchmarks of empirical calibration software packages. Run
  time in milliseconds.} \label{table:benchmark}
\end{table}

\section{Conclusion}

Empirical calibration has many applications, particularly in survey sampling and
causal inference. \pkg{EC} provides an easy to use, robust, and efficient
implementation. The source code is shared at
\url{https://github.com/google/empirical_calibration}, where additional
documentation and application examples can be found.

\subsubsection*{Acknowledgments}

The authors thank Art Owen, Jim Koehler, Joseph Kelly, Georg Goerg, Jon Vaver,
Susanna Makela, Mike Hankin, and Mike Wurm for helpful discussions. The authors
appreciate Tony Fagan, Penny Chu, and Elissa Lee for their support. Special
thanks go to Jim Koehler, David Chan, and Tim Au for reviewing this manuscript.

\bibliography{references}
\bibliographystyle{iclr2019_conference}

\clearpage

\begin{appendices}

\section{Convex Optimization Details}
\label{app:cvx}

\subsection{Solving the Dual Problem}

We present a detailed solution for the quadratic loss. The solution for entropy
loss can be derived similarly. We first convert the optimization into vector
form
\begin{align}
\underset{\rvw}{\text{minimize}} \quad
& \frac{1}{2} \rvw^T \rvw \label{quadratic-objective} \\
\text{subject to} \quad
& Z^T \rvw = \mathbf{0}_p, \label{hard-constraint} \\
& \mathbf{1}_n^T \rvw = 1, \label{weight-normalization} \\
& \rvw \ge \mathbf{0}_n,
\end{align}
where $Z$ is the $n \times p$ covariate matrix subtracted by the marginal
constraint, i.e., the $i$-th row $Z_i = \rvc(X_i) - \mathbf{\bar c}$.

We construct the Lagrangian with $\beta \in \mathbb{R}^p$, $\theta \in
\mathbb{R}$, and $\alpha \in \mathbb{R}_{\ge 0}^n$ as the Lagrangian multipliers
\begin{equation*}
L =\frac{1}{2} \rvw^T \rvw −\beta^T Z^T \rvw − \theta (\mathbf{1}_n^T \rvw − 1)
− \alpha^T \rvw.
\end{equation*}

Setting $\frac{\partial L}{\partial w_i}$ to zero reveals that the primal
weight is a linear function of the transformed covariate and Lagrangian multipliers
\begin{equation*}
w_i = Z_i^T \beta + \theta + \alpha_i.
\end{equation*}

Multiplying $w_i$ on both sides and using the complementary slackness
Karush-Kuhn-Tucker condition $w_i \alpha_i = 0$, we eliminate $\alpha_i$ and
obtain
\begin{equation}
w_i (w_i − Z_i^T \beta − \theta) = 0,
\end{equation}
which further reduces to
\begin{equation} \label{quadratic-link}
w_i =\max \{0,\ Z_i^T \beta + \theta\}.
\end{equation}

Substituting~\eqref{quadratic-link} into
~\eqref{hard-constraint}-\eqref{weight-normalization}, we obtain a
$p+1$-dimensional estimating equation with respect to the dual parameter
$(\beta,\ \theta)$. The equation can be solved using a general purpose nonlinear
equation solver, such as
\href{https://docs.scipy.org/doc/scipy/reference/generated/scipy.optimize.fsolve.html}{\texttt{scipy.optimize.fsolve}}.
When such a solution exists, it would be unique because the
quadratic objective is strictly convex.

The dual solution for the entropy objective can be derived similarly, but with
an exponential link function to replace~\eqref{quadratic-link}:
\begin{equation}
w_i = e^{Z_i^T \beta + \theta} \label{entropy-link}.
\end{equation}

By construction of~\eqref{quadratic-link} and~\eqref{entropy-link}, sample data
points with identical covariates would be given identical weights, which is a
desirable property.

\subsection{Bounding the Weights}

For certain applications, it is desirable to impose additional upper and/or
lower bound on the weights to reduce the impact of extreme weights. Suppose the
imposing bound is $[l, u]$, where $l$ and $u$ are both non-negative. We only
need to change the weight link functions accordingly. \eqref{quadratic-link}
would become
\begin{equation*}
w_i =\min \{u,\ \max \{l,\ Z_i^T \beta + \theta\}\},
\end{equation*}
and~\eqref{entropy-link} is now
\begin{equation*}
w_i =\min \{u,\ \max \{l,\ e^{Z_i^T \beta + \theta}\}\}.
\end{equation*}

\subsection{Relaxing the Equality Constraint}

When there is no feasible solution, we need to either drop some marginal
constraints or soften the hard constraints as in soft margin support vector
machine
\begin{equation} \label{soft-constraint}
 | Z^T \rvw |_2 \le \epsilon,
\end{equation}

where $\epsilon$ is the tolerance. It only requires slightly more work to solve
optimization problem with tolerance added. Following the common practice, we
first introduce two $p$-dimensional non-negative slack variables $\Delta_1$ and
$\Delta_2$, and reformulate the optimization problem as
\begin{align}
\underset{\rvw}{\text{minimize}} \quad
& \frac{1}{2} \rvw^T \rvw \\
\text{subject to} \quad
& Z^T \rvw + \Delta_1 − \Delta_2 = \mathbf{0}_p,
\label{soft-constraint-slack} \\
& \mathbf{1}_n^T \rvw = 1, \label{weight-normalization2} \\
& \rvw \ge \mathbf{0}_n, \\
& \epsilon^2 − \Delta_1^T \Delta_1 − \Delta_2^T \Delta_2 \ge 0, \\
& \Delta_1 \ge \mathbf{0}_p, \\
& \Delta_2 \ge \mathbf{0}_p.
\end{align}

We then construct the Lagrangian with $\beta \in \mathbb{R}^p$, $\theta \in
\mathbb{R}$, $\alpha \in \mathbb{R}_{\ge 0}^n$, $\lambda \in \mathbb{R}_{\ge
  0}$, and $\gamma_1,\ \gamma_2 \in \mathbb{R}_{\ge 0}^p$ as the Lagrangian
multipliers
\begin{equation*}
L = \frac{1}{2} \rvw^T \rvw − \beta^T (Z^T \rvw + \Delta_1 − \Delta_2) − \theta
(\mathbf{1}_n^T \rvw − 1) − \alpha^T \rvw − \lambda (\frac{1}{2} \epsilon^2 −
\frac{1}{2} \Delta_1^T \Delta_1 − \frac{1}{2} \Delta_2^T \Delta_2 ) − \gamma_1^T
\Delta_1 − \gamma_2^T \Delta_2.
\end{equation*}

Setting $\frac{\partial L}{\partial w_i}$ to zero and applying the
complementary slackness KKT condition $w_i \alpha_i = 0$ leads to the same
solution~\eqref{quadratic-link} for $w_i$.

Setting $\frac{\partial L}{\partial \Delta_1}$ and $\frac{\partial L}{\partial
  \Delta_2}$ to zero and applying their respective complementary slackness KKT
conditions, we have
\begin{align*}
\lambda \Delta_1 & = \max\{0,\ \beta\}, \\
\lambda \Delta_2 & = \max\{0,\ −\beta\},
\end{align*}
which in turn leads to
\begin{align}
\lambda (\Delta_1 − \Delta_2) & = \beta, \label{slack-diff} \\
\lambda^2 (\Delta_1^T \Delta_1 + \Delta_2^T \Delta_2) & =
 |\beta |_2^2. \label{slack-norm}
\end{align}

As a result the slack variables can be eliminated
from~\eqref{soft-constraint-slack}
\begin{equation} \label{soft-constraint-lambda}
\lambda Z^T \rvw + \beta = \mathbf{0}_p.
\end{equation}

Combining~\eqref{slack-norm} and the complementary KKT condition
\begin{equation*}
\lambda
(\epsilon^2 − \Delta_1^T \Delta_1 − \Delta_2^T \Delta_2) = 0,
\end{equation*}
we get
\begin{equation*}
\lambda \epsilon = |\beta |_2,
\end{equation*}
which nicely connects the $L_2$ tolerance $\epsilon$ to $L_2$ norm of the dual
parameter $\beta$, and further eliminates $\lambda$
from~\eqref{soft-constraint-lambda},
\begin{equation} \label{soft-constraint-epsilon}
Z^T \rvw + \frac{\epsilon \beta}{|\beta |_2} = \mathbf{0}_p.
\end{equation}

It reveals that it is the scaled dual parameter $\beta$ that serves as the
\emph{slack} for the relaxed covariate balance
constraint~\eqref{soft-constraint-slack}. Finally,
combining~\eqref{quadratic-link}, \eqref{weight-normalization2}
and~\eqref{soft-constraint-epsilon}, we again obtain a $p+1$-dimensional
estimating equation with respect to the dual parameter $(\beta,\ \theta)$, and
can solve it using a general purpose nonlinear equation solver.

\section{Empirical Calibration and Other Causal Methods}
\label{app:causal}

We briefly compare empirical calibration with other causal inference
methods. Randomized experiments use a randomized assignment mechanism to ensure
that the covariate distributions between the treated and control groups are
naturally balanced. For observational data, matching and weighting are the two
main methods to enforce this balance. \textit{Matching} refers to matching
treatment and control units at the individual level; while \textit{weighting}
assigns continuous weights to control units such that their weighted average
matches the treatment average. \citet{zhao2017entropy} nicely summarized all
matching and weighting methods in the table below

\begin{table}[ht]
\centering
\begin{tabular}[t]{lrr}
  \toprule
  & Discrete weights & Continuous weights \\
  \midrule
  By raw covariates & Raw matching & Empirical calibration \\
  By propensity scores & Propensity score matching & Propensity score weighting \\
  \bottomrule
\end{tabular}
\caption{Causal methods with observational data.}
\end{table}

Weighting is generally preferred to matching due to its efficiency gain
\citep{zhao2017entropy}. Compared to propensity score weighting, empirical
calibration enjoys similar theoretical properties but yields more stable
weights.

\paragraph{Raw Matching} uses raw pre-treatment covariates for matching and does
not attempt to model the assignment mechanism. Both exact matching and $k:1$ nearest
neighbor matching fall into this category. Their variants vary by
choice of the distance metric, and whether the same control unit is allowed to be
used multiple times as a match (matching with replacement).

Raw matching is intuitive and easy to implement. In many applications, however,
it can be difficult or impossible to find $k$ closely matched control units for
certain treatment units. Empirical calibration, on the other hand, aims to
balance the treated and control group at the aggregate level, which is usually
sufficient for estimating ATT.

Nearest neighbor matching implies binary or discrete weights to control
units. For an observational data with a large control group, many control units
would be effectively discarded in the analysis. By adopting continuous weights,
empirical calibration uses information from more control units. It leads to a
larger effective sample size for the control group and reduces the variance of
the ATT estimate.

\paragraph{Propensity Score Matching (PSM)} assumes a regression model for the
propensity score $p(X) = P(T=1 | X)$. The treated and control units are matched
using the estimated propensity scores. The propensity score serves as the
\textit{sufficient statistic} of treatment assignment, and reduces the problem
to a univariate matching problem.

The success of the propensity score method hinges on the quality of the estimated
scores. Accurately estimated propensity scores stochastically balance the
covariates between the treatment and control groups. In practice, however,
matching or weighting solely based on propensity score rarely guarantees actual
covariate balancing. If the data is balanced to begin with, PSM may
increase the imbalance \citep{king2015propensity}. Empirical calibration takes the
guesswork out of the procedure by directly optimizing towards an explicit
covariate balancing goal, with guaranteed results.

\paragraph{Propensity Score Weighting (PSW)} weights the control units using the
\textit{inverse probability weighting} (IPW) scheme
\begin{equation}
\hat \tau_{\text{IPW}} = \frac{1}{n_1} \sum_{i: T_i=1} Y_i -
\frac{1}{n_0}\sum_{i: T_i=0} \frac{\hat p(X_i)}{1 - \hat p(X_i)} Y_i,
\end{equation}
where $\hat p(X_i)$ is the estimated propensity score for the $i$-th unit.

The inverse probability weights could be volatile and sensitive to model
misspecification. If some estimated propensity score is close to zero for a
control unit, its inverse weight becomes very large and unstable, inflating the
variance of the IPW estimate for ATT. One remedy is to trim extreme weights or
apply winsorization, but it is hard to choose the trimming threshold in a
principled way. Empirical calibration seeks weights that least deviate from
uniform weights, so large weights are naturally penalized in the optimization.
The resulting weights are far more stable. Furthermore, if desired, bounds on
the weights can be directly imposed in optimization without the need for a
separate trimming step.

Conventionally, propensity score regression and outcome regress models are
fitted separately and then combined to construct a \textit{doubly robust}
estimator --- if either model is correctly specified, the estimator is
statistically consistent. Recently, \citet{zhao2017entropy} showed that entropy
balancing (EB) is also doubly robust with respect to the implied logistic
propensity score regression and linear outcome regression.

\citet{hirano2003efficient} showed that IPW paired with the sieve propensity score
model could achieve the semiparametric efficiency bound, which justifies why
weighting is usually preferred over matching. Empirical calibration with the
entropy loss has also been shown to reach the asymptotic semiparametric
efficiency bound when both implied propensity score regression and outcome
regression are correctly specified \citep{zhao2017entropy}.

\end{appendices}

\end{document}